\documentclass[useAMS,usenatbib]{mn2e}
\usepackage{graphicx}
\usepackage{epsfig}

\begin{document}
\title[IGR J17544--2619]{Discovery of the orbital period in the supergiant fast X-ray transient IGR J17544--2619}
\author[D. J. Clark et al.] {D. J. Clark$^{1}$\thanks{E-mail: djc400@soton.ac.uk}, A. B. Hill$^{2}$, A.J Bird$^{1}$, V. A. McBride$^{1}$, S. Scaringi$^{1}$, A.J Dean$^{1}$\\
$^{1}$School of Physics and Astronomy, University of Southampton, Southampton. SO17 1BJ, UK\\
$^{2}$Laboratoire d'Astrophysique de Grenoble, UMR 5571 CNRS, Universit\'e Joseph Fourier, BP 53, 38041 Grenoble, France}

\date{Accepted 2009 August 5.  Received 2009 August 5; in original form 2009 July 16}

\pagerange{\pageref{firstpage}--\pageref{lastpage}} \pubyear{2009}

\maketitle
\label{firstpage}
\begin{abstract}
The supergiant fast X-ray transient (SFXT) system IGR J17544--2619 has displayed many large outbursts in the past and is considered an archetypal example of SFXTs.  A search of the {\it INTEGRAL/ISGRI} data archive from MJD 52698--54354 has revealed 11 outbursts and timing analysis of the light curve identifies a period of 4.926$\pm$0.001 days which we interpret as the orbital period of the system. We find that large outbursts occasionally occur outside of periastron and place an upper limit for the radius of the supergiant of $<$23R$_{\sun}$.
\end{abstract}
\begin{keywords}
Gamma-rays: observations - X-rays: binaries - X-rays: individual: IGR J17544-2619
\end{keywords}
\section{Introduction}
Discovered on the 17th September 2003 by the {\it IBIS/ISGRI} instrument \citep{Ubertini:2003kn, 2003A&A...411L.141L} onboard the  {\it INTEGRAL} satellite \citep{Winkler:2003hh}, IGR J17544--2619 has become the archetypal example of the Supergiant Fast X-ray Transient (SFXT) class of objects \citep{Sguera:2005os,Negueruela:2006lt}.  Around twenty SFXTs are known to date and as the name suggests this class of object is characterized by rapid outbursts, lasting typically a few hours, with a flux 10$^3$--10$^4$ times the faint quiescent emission X-ray luminosities in the range of 10$^{32}$--10$^{34}$ ergs s$^{-1}$.  Through analysis of optical and NIR spectra and periodicities in the X-ray light curves, SFXTs have been shown to consist of a supergiant donor and a compact accreter.  The observation of spin periods in a few SFXTs (IGR J16465--4507, AX J1841.0--0535 and IGR J18483--0311) show that the compact object is often a neutron star.

A number of hypotheses to explain the flaring behavior of SFXTs have been proposed.  A common interpretation invokes a structured wind from the supergiant companion, either in the form of clumping of a spherically symmetric outflow from the donor star \citep{Zand:2005dp,Walter:2007ww, Leyder:2007uo, Ducci:2009yq}, or a wind with an equatorial density profile inclined to the compact object orbit \citep{Sidoli:2007ek}.  In either case variations in the orbital eccentricity \citep{2008AIPC.1010..252N} can enhance the transient nature of the emission from these supergiant binary systems.  A gated accretion has also been proposed \citep{2008ApJ...683.1031B} where transitions between different accretion regimes due to changes in the magnetospheric radius cause step-like changes in the luminosity. 

The initial discovery  of IGR J17544--2619 on the 17th September 2003 was triggered by the {\it INTEGRAL} observation of a 2\,h outburst and an 8\,h outburst on the same day, with a further outburst of 10\,h detected on 8th March 2004 demonstrating that this was a fast transient source.  Subsequently, more bursts from the source were also discovered in older {\it BeppoSAX} data \citep{Zand:2004ge} and recently seen in {\it XMM-Newton} \citep{2004AA...420..589G}, {\it Chandra} \citep{Zand:2005dp}, {\it Swift} \citep{Krimm:2007ws, Sidoli:2009kl, Krimm:2009ht} and {\it Suzaku} data \citep{Rampy:2009tt}. Due to observation strategies and the relatively rare occurrences only a few bursts are observed a year and therefore identification of any binary system parameters is challenging.

IGR J17544--2619 is situated $\sim$3$^\circ$ from the Galactic centre ({\it l} = 3.24$^\circ$, {\it b} = --0.34$^\circ$), a region shown to contain many recurrent transient sources \citep{Kuulkers:2007um}.  The location of IGR J17544--2619 is in a well studied region covered by a large amount of {\it INTEGRAL} archive data and is now covered by the Galactic Center Bulge Monitoring campaign.  In this paper we report on the detection of a modulation of the gamma-ray flux interpreted as the orbital motion and report new outbursts discovered by a search of the {\it INTEGRAL} light curve.
\section{Previous Observations}

The first reported detection of IGR J17544--2619 was on 17th September 2003  \citep{Sunyaev:2003xj} when two flares reaching a maximum of 160\,mCrab (18-25 keV) were detected by {\it IBIS/ISGRI} \citep{Sunyaev:2003xj,Grebenev:2003ez}.  A further 5 bursts were seen when looking back at the older {\it BeppoSAX}-WFC data (\citealt{Zand:2004ge}, see Table~\ref{table:oldbursts}) and another from 6\,d before the {\it INTEGRAL} observation in {\it XMM-Newton} data \citep{2004AA...420..589G}.  In total 8 bursts have been previously reportedly in the {\it INTEGRAL} data \citep{Walter:2007ww}, although the exact times have not been published for all the bursts.  

\begin{table*}
	\centering
	\caption{Bursts found in literature and in this paper. The peak flux is the 18--60\,keV maximum average science window flux, assuming an N$_H = 3.8\times10^{22}$ and power law index $\alpha$ = 2.25 \citep{2004AA...420..589G}. $\star$ represents bursts found in our light curve. (1) not seen in {\it INTEGRAL} light curve due to observation gap. ${^\dagger}$ 0.5-10\,keV.}
	\begin{tabular}{lcr@{.}lr@{.}lr@{.}lccl}
	\hline
	\multicolumn{1}{c}{Date} & Orbital & \multicolumn{2}{c}{Duration} 	& \multicolumn{2}{c}{Sig} 	& \multicolumn{2}{c}{Peak Flux } & \multicolumn{1}{c}{Instrument} 		& \multicolumn{1}{c}{Reference} 	\\
	\multicolumn{1}{c}{MJD} & Phase & \multicolumn{2}{c}{Hours}	& \multicolumn{2}{c}{$\sigma$}			& \multicolumn{2}{c}{$10^{11}$erg\,cm$^{-2}$s$^{-1}$}&  								&           		\\
	\hline
    	50320.6 	& 0.39 &	3&3				& \multicolumn{2}{c}{}	& 	\multicolumn{2}{c}{}	& SAX/WFC 		& \citet{Zand:2004ge} 		& ¥ \\
    	51229.7 	& 0.94 &	0&2 				& \multicolumn{2}{c}{}	& 	\multicolumn{2}{c}{}	& SAX/WFC 		& \citet{Zand:2004ge} 		& ¥ \\
    	51248.8 	& 0.81 &	0&4 				& \multicolumn{2}{c}{}	& 	\multicolumn{2}{c}{}	& SAX/WFC 		& \citet{Zand:2004ge} 		& ¥ \\
    	51807.5 	& 0.23 &	8&3 				& \multicolumn{2}{c}{}	& 	\multicolumn{2}{c}{}	& SAX/WFC 		& \citet{Zand:2004ge} 		& ¥ \\
    	51825.1 	& 0.80 &	1&0    			& \multicolumn{2}{c}{}	& 	\multicolumn{2}{c}{}	& SAX/WFC 		& \citet{Zand:2004ge} 		& ¥ \\
	52893.8	& 0.75 &	$\sim$2&0		& \multicolumn{2}{c}{}	&	~~~~~~~4&0$ {^\dagger}$ 	& XMM-Newton	& \citet{2004AA...420..589G}	& \\
    	52899.05 	& 0.81 &	12&94 			&     	~33&24 			&     	111&2$\pm$4		& INTEGRAL/ISGRI & \citet{Sunyaev:2003xj}		& $\star$ \\
    	53062.59 	& 0.01 &	1&12 			&      5&76 			&    	28&6$\pm$6		& INTEGRAL/ISGRI & ¥ 						& $\star$ \\
    	53072.35 	& 0.99 &   5&34 			&    	30&13 			&     	119&7$\pm$6		& INTEGRAL/ISGRI & \citet{Grebenev:2004iw}		& $\star$ \\
    	53189	& 0.67 &	0&8				& \multicolumn{2}{c}{}	&	230&0$ {^\dagger}$ & Chandra		& \citet{Zand:2005dp}		& (1) \\
    	53269.88 	& 0.09 &   4&22 			&     	10&41 			&    	39&$\pm$11		& INTEGRAL/ISGRI & \citet{Sguera:2006ns}		& $\star$ \\
    	53441.23 	& 0.88 &   0&56 			&     	10&59 			&   	62&6$\pm$6 		& INTEGRAL/ISGRI & \citet{Sguera:2006ns} 		& $\star$ \\
    	53481.26 	& 0.00 &   1&88 			&     	12&20 			&     	42&1$\pm$5 		& INTEGRAL/ISGRI & ¥ 						& $\star$ \\
    	53628.67 	& 0.93 &   5&06 			&   	6&13 			&    	44&1$\pm$9 		& INTEGRAL/ISGRI & ¥ 						& $\star$ \\
    	53656.92 	& 0.66 & 	34&69 			&     	12&99 			&   	73&5$\pm$6 		& INTEGRAL/ISGRI & ¥ 						& $\star$ \\
    	53806.87 	& 0.10 &   3&75 			&     	16&91 			&   	70&4$\pm$7 		& INTEGRAL/ISGRI & ¥ 						& $\star$ \\
    	53987.52 	& 0.77 &   2&06 			&    	8&32 			&    	34&5$\pm$5 		& INTEGRAL/ISGRI & ¥ 						& $\star$ \\
    	53998.38 	& 0.98 &   4&59 			&     	12&61 			&   	102&4$\pm$17 	& INTEGRAL/ISGRI & ¥ 						& $\star$ \\
    	54364.3 	& 0.26 & 	\multicolumn{2}{c}{}	& \multicolumn{2}{c}{} 	& 	\multicolumn{2}{c}{}	& INTEGRAL/ISGRI 	& \citet{Kuulkers:2007um}	& ¥ \\
    	54412.1 	& 0.96 &	\multicolumn{2}{c}{}	& \multicolumn{2}{c}{} 	& 	\multicolumn{2}{c}{} 	& Swift 			& \citet{Krimm:2007ws} 		& ¥ \\
    	54556.9 	& 0.35 &	0&13 			& \multicolumn{2}{c}{}	& 	\multicolumn{2}{c}{}	& Swift, Suzaku 	&\citet{Sidoli:2009kl}, \citet{Rampy:2009tt}& ¥ \\
    	54570.4 	& 0.10 &	\multicolumn{2}{c}{}	& \multicolumn{2}{c}{}	& 	\multicolumn{2}{c}{} 	& INTEGRAL/ISGRI 	& \citet{Kuulkers:2007um}	& ¥ \\
    	54708.1 	& 0.05 & 	\multicolumn{2}{c}{}	& \multicolumn{2}{c}{}	& 	\multicolumn{2}{c}{} 	& INTEGRAL/ISGRI 	& \citet{Kuulkers:2007um} 	& ¥ \\
    	54713    	& 0.05 & 	\multicolumn{2}{c}{}	& \multicolumn{2}{c}{} 	& 	\multicolumn{2}{c}{} 	& Swift 			& \citet{Sidoli:2009kl}		& ¥ \\
    	54905    	& 0.02 & 	$\sim$0&55		& \multicolumn{2}{c}{}	& 	\multicolumn{2}{c}{} 	& Swift 			&  \citet{Krimm:2009ht}		& ¥ \\
	\hline
	\end{tabular}
	\label{table:oldbursts}
\end{table*}

Observations using {\it XMM-Newton} and {\it Chandra} provided precise positional errors of 4" and 0.6" respectively supporting the optical/NIR counterpart 2MASS J17542527--2619526 \citep{Rodriguez:2003af} over the {\it ROSAT} candidate 1RXS J175428.3--2620 \citep{Wijnands:2003ow}.  Using combined optical and X-ray observations \citet{Pellizza:2006ia} classified the counterpart of IGR J17544--2619 to be a O9Ib with a mass of 25--28M$_{\sun}$. Using optical and NIR data, \citet{Pellizza:2006ia} estimated a distance of 2.1--4.2\,kpc based on extinction measurements.  \citet{Rahoui:2008ig} used the data from \citet{Pellizza:2006ia} to fit the SED and obtain a radius range for the donor star of 12.7R$_{\sun}<R_{\star}<26.6R_{\sun}$. 

An analysis of {\it Chandra} data during quiescence by \citet{Zand:2005dp} produced a power law fit with $\Gamma = 5.9\pm1.2$, much softer than that expected for a black hole, suggesting that the compact object was a neutron star. If the accretor is a black hole then a radio flux of $\sim$10-35\,mJy is expected.  However, an upper limit of 7\,mJy has been placed on the radio emission of IGR J17544--2619 adding to the evidence that the compact object is a neutron star \citep{Pellizza:2006ia}.  In the analysis of {\it Suzaku} observations, \citet{Rampy:2009tt} found that the absorption showed varied rapidly on timescales of minutes by  a factor of $\sim$2 during flaring, interpreted as a wind clump passing though the line of sight.  However, longer integrations showed stable values of N$_H$ compatible with the stable absorption seen by \citet{Zand:2005dp}, but this may simply be the result of averaging.

The initial bursts found by {\it INTEGRAL} led to a prediction of a 165$\pm$3 day period in the source \citep{Walter:2006fv}.  This period was supported by the few bursts found in  {\it BeppoSAX} and {\it Swift} data.  Long period systems can be difficult to explain \citep{1995MNRAS.274..461B}.  A long period requires a large eccentricity to explain any substantial accretion, but this would decay quickly and require fine tuning not to disrupt the binary.  Therefore, it would be unlikely to see too many of these systems.  No evidence for spin or orbital periods has been observed in {\it XMM-Newton} or {\it Chandra} data.
\section{Data Analysis}
Archival IBIS data from MJD 52698 to MJD 54354, giving an on source exposure of $\sim8$\,Msec, were processed with the {\it INTEGRAL} Off-line Science Analysis (OSA \citealt{2003A&A...411L.223G}) version 7.0.  A light curve for IGR~J17544$-$2619 was generated on science window timescales ($\sim2000$\,s) over the 18-60\,keV energy range.  The light curve was then searched for bursts.  The shortest timescale region of the light curve that produces the highest detection significance greater than 4\,$\sigma$ is found.  This is the brightest burst in the light curve.  This section of the light curve is then removed from the total light curve and the next highest significant section is searched for. 
\begin{figure}
	\includegraphics[width=1.0\linewidth,trim = 0mm 0mm 0mm 6mm, clip]{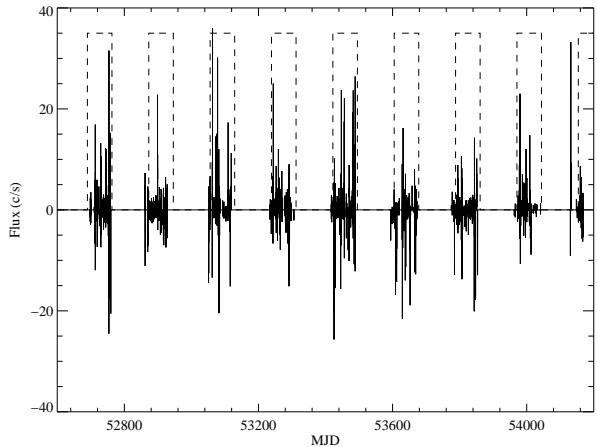}
	\caption{Overlaying a 185 day 40\% duty cycle period on the 18--60\, keV {\it ISGRI} science window light curve clearly shows that this period is a result of the {\it INTEGRAL} observing strategy.}
	\label{figure:lcphase}
\end{figure}
In total 11 bursts have been found in the light curve (Table~\ref{table:oldbursts}) which can be added to the list of bursts found in the literature.  The bursts range from short flares of 0.5\,h to periods of flaring activity up to 35 hours. The time between bursts ranges from less than 9 days to 197 days.  By folding the burst times on different periods, an estimation of the duty cycle required over the orbit to explain all the bursts can be made.  There is a clear minimization of $\sim$40\% duty cycle at  a period of 185 days in our data.  However, this period can be shown to be a result of the observing strategy for the source (Fig.~\ref{figure:lcphase}).  {\it INTEGRAL} is restricted in observing the Galactic center due to visibility constraints including the 40$^{\circ}$ solar aspect angle due to the fixed solar arrays and so there is an $\sim$6 month period to the observations of IGR J17544--2619.  Overlaying this period and duty cycle on the {\it ISGRI}  light curve, it is clear that this period is a result of the observing strategy, and with only a few points of reference this could be mistaken for a 165 day period suggested by \citet{Walter:2006fv}. No other significant periods were found by the initial folding analysis.
\section{Periodicity Analysis}
\label{section:period}
\begin{figure}
\centering
\includegraphics[width=1.0\linewidth,trim = 0mm 0mm 0mm 6mm, clip]{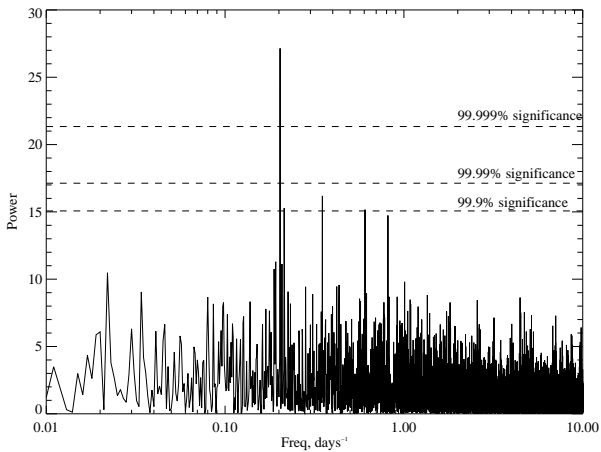}
\caption{The Lomb-Scargle periodogram of the IBIS/ISGRI 18--60 keV light curve.  The 99.9\%, 99.99\% and 99.999\% significance levels derived from randomization tests are over-plotted.}
\label{fig:LS}
\end{figure}

The science window light curve (containing 6443  independent data points) was searched for signs of periodicity using the Lomb-Scargle periodogram method \citep{1976Ap&SS..39..447L, 1982ApJ...263..835S, 1989ApJ...343..874S}.  The Lomb-Scargle periodogram is shown in Figure~\ref{fig:LS} with a clear peak evident at a frequency of 0.203 days$^{-1}$, corresponding to a period of 4.926 days.  The significance of the peak was confirmed by performing a randomization test \citep{Hill:2005jy}; the flux points of the light curve are randomly re-ordered and a new Lomb-Scargle periodogram generated, the distribution of the power of the largest peak is an indicator of the significance of detection.  From 200,000 simulations we estimate the 99.9\%, 99.99\% and the 99.999\% significance levels which are shown in Figure~\ref{fig:LS}; note that the maximum power achieved in any of the randomized light curves was $<$22.

The error on the identified period was estimated using a Monte-Carlo simulation. Each flux measurement was adjusted using Gaussian statistics within its individual error estimate to generate a simulated light curve of the source. The corresponding periodogram was produced and the location of the maximum peak near the frequency of 0.203 days$^{-1}$ was recorded. From 200,000 simulations we estimate the period and its error to be 4.926$\pm$0.001 (1$\sigma$) days.

\begin{figure}
\centering
\includegraphics[width=1.0\linewidth,trim = 0mm 0mm 0mm 6mm, clip]{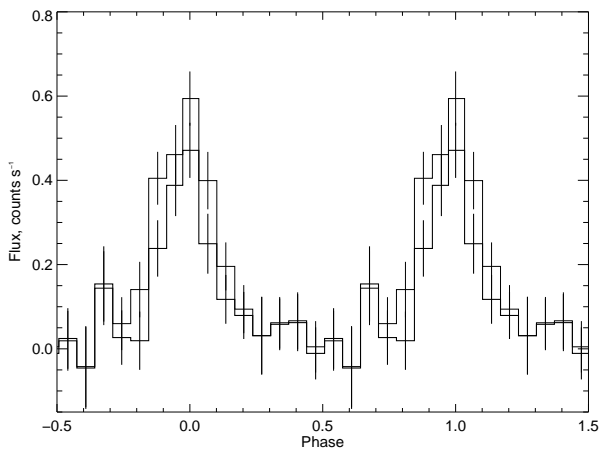}
\caption{The IBIS/ISGRI 18--60 keV light curve of IGR J17544$-$2619 folded on 4.926 days and using an ephemeris of 52702.9 days.  The narrow line is for the whole data set, while the bold line is the result when all of the outbursts are removed.}
\label{fig:fold}
\end{figure}

Figure~\ref{fig:fold} shows the phase folded light curve using a period of 4.926 days and using an ephemeris of MJD 52702.9 to centre the fold at phase 0.0.  Since we would expect any enhancement to the emission to coincide with the neutron stars closest approach to the donor, we consider phase 0.0 to be periastron for the rest of the paper. Between phases 0.25 and 0.85 the folded light curve is consistent with zero which could be interpreted as an eclipse.  However, this eclipse would last ~3\,d, ~60\% of the orbit and so is unlikely.  To see the effect of the outbursts on the phase folded curve a new fold was performed excluding all of the outbursts detected (Table~\ref{table:oldbursts}).  This second phase folded light curve is also shown in Figure~\ref{fig:fold}, and shows no substantial difference in shape, demonstrating that there is an underlying periodicity in the flux aside from the bursts when then entire light curve is considered.  The public light curves for IGR J1755$-$2619 from the \emph{RXTE}-ASM and \emph{Swift}-BAT missions were also analyzed using the Lomb-Scargle periodogram to see if there was any indication of the 4.926 day periodicity.  Neither data set showed an indication of a periodic signal. Folding the \emph{RXTE}-ASM and \emph{Swift}-BAT light curves on the {\it INTEGRAL} period similarly showed no indication of a modulation.

\section{Outburst Recurrence}
Aside from the bursts that have been reported in Table~~\ref{table:oldbursts}, there is the possibility of lower level emission occurring at each periastron pass that does not show up as a burst during our search.  In order to test the recurrence of emission at periastron we carry out a recurrence analysis \citep{Bird:2009yq}.  For this test we calculate the light curve significance for a region around the orbital period and at the midpoint of the period.  This will then show the recurrence of bursts at periastron and if we are missing any low level emission seen at every periastron.

\begin{figure}
	\includegraphics[width=1.0\linewidth,trim = 0mm 0mm 0mm 6mm, clip]{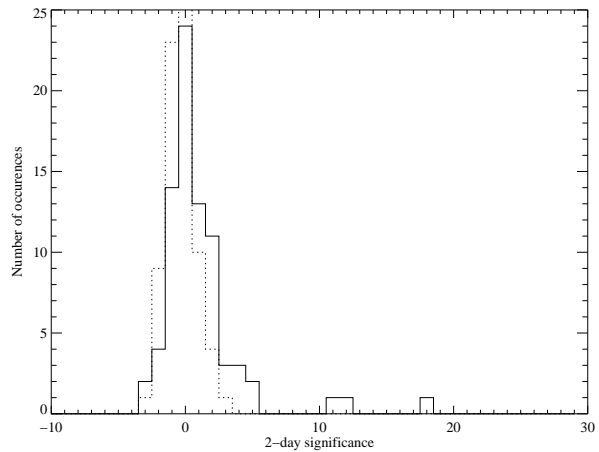}
 	\caption{Recurrence analysis of IGR J17544--2619 light curve.  A comparison of the 2-day significance at periastron (solid line) and the 2-day significance and apastron (dotted line) shows that the only difference is due to the bursts.}
	\label{figure:recurrence}
\end{figure}

In the case of IGR J17544--2619 the bursts do not all occur at periastron, reducing the sensitivity of this test.  Since this orbit is short ($\sim$5 days) we take a region of 1 day either side of periastron (Fig.~\ref{figure:recurrence}). Comparing the distributions of significances for apastron and periastron using a Kolmogorov-Smirnov test we find them to be different at the 99\% level.  Out of the 79 periastrons seen only 11 produce a significance over 3$\sigma$.  The difference between the distributions at periastron and apastron can be explained by the bursts seen and not an underlying longer timescale variation as seen in SAX J1818.6--1703 or the {\it Suzaku} data for IGR J17544--2619.  If we repeat this with the known bursts removed the KS test finds that the two distributions are not significantly different demonstrating the underlying emission, which can be seen with all the phase data added together in the Lomb-Scargle analysis, cannot be seen when considering individual periastrons.
\vspace{-12pt}
\section{Discussion}
 This is the third SFTX with a period of a few days. The other two having very similar properties to IGR J17544-2619. IGR J16418-4532 \citep{2006ATel..779....1C} and IGR J16479-4514 \citep{2009MNRAS.397L..11J} have periods of 3.75\,d and 3.32\,d respectively. This contradicts the view that SFXTs have different orbital geometries to the classical HMXBs \citep{Walter:2007ww} and suggests that a difference in the stellar wind may lead to the difference between SFXTs and HMXBs.

\begin{figure}
	\includegraphics[width=1.0\linewidth,trim = 0mm 0mm 0mm 6mm, clip]{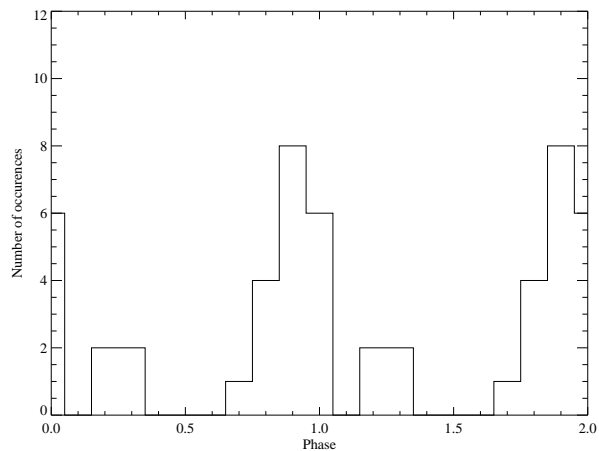}
 	\caption{Folding the bursts on 4.926 days shows that burst may occur at any time during the orbit, but are more probable at phase 1.0, likely to be periastron.}
	\label{figure:burstphase}
\end{figure}

Using the ephemeris calculated by the timing analysis, a histogram of burst times with respect to orbital phase can be produced (Figure~\ref{figure:burstphase}).  This shows while bursts may occur at any time during the orbital period they are most likely in the last 20\% of the orbit, suggesting that this phase is the point of closest approach of the orbit between the two bodies. 
\begin{figure}
	\includegraphics[width=1.0\linewidth,trim = 0mm 0mm 0mm 6mm, clip]{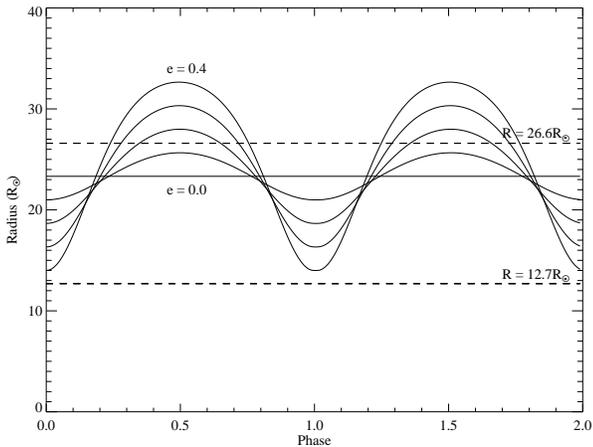}
	\caption{The Lagrange point radius between the neutron star and supergiant as a function of phase for different eccentricities of orbit ({\it e} = 0, 0.1, 0.2, 0.3, 0.4 solid lines).  If the Lagrange point is less than the supergiant radius (dashed lines are the upper/lower limits) then Roche-lobe overflow is expected \citep{1971ARA&A...9..183P}.}
	\label{figure:roche}
\end{figure}
Since we have a mass for the supergiant of 25-28M$_{\sun}$ and assuming a mass of  1.4 M$_{\sun}$ for the compact object, which is probably a neutron star, Keplers 3$^{rd}$ law gives a semi-major axis of 36-38 R$_{\sun}$ for the 4.9d period.  The range of possible radii for an O9Ib star are 12.7-26.6R$_{\sun}$ \citep{Pellizza:2007xa, Rahoui:2008ig}.

Figure~\ref{figure:roche} shows how the Lagrange point between the two stars changes with orbit for different eccentricities \citep{1971ARA&A...9..183P}. We do not see any persistent or regular emission every orbit and so there should be no Roche-lobe overflow in the system. This limits the size of the star to $<$23R$_{\sun}$ with no eccentricity in the orbit. If the minimum possible radius of the star is used then this limits the eccentricity of the orbit to less than $<$0.4. 

In SAX J1818.6-1703, \citet{2009A&A...493L...1Z} proposed that the wind and compact object velocities will limit the region over which the NS can accrete to less than a few stellar radii.  With the short period detected in IGR J17544--2619 the neutron star will stay relatively close to the donor star.  \citet{2009A&A...493L...1Z} describe how the accretion rate will vary strongly with distance from the supergiant, which may be the cause of the underlying modulation seen in the folded light curve (Fig.~\ref{fig:fold}). With bursts seen at all phases this suggests the neutron star is within this accretion envelope and the variation could be due to the change in wind/NS velocity. This variation could also be due to eccentricity effecting the probability of seeing a burst.

\begin{figure}
	\includegraphics[width=1.0\linewidth,trim = 0mm 0mm 0mm 6mm, clip]{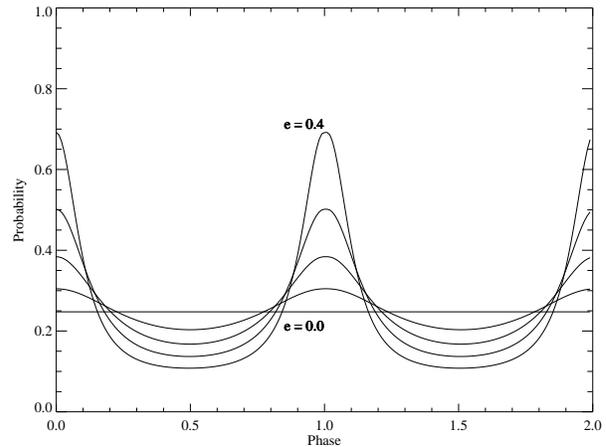}
	\caption{If the probability of a stellar wind clump hitting the neutron star is proportional to a  $(r^{2/3}-r^{1/3})/r^2$ law then even a slight eccentricity ({\it e} = 0, 0.1, 0.2, 0.3, 0.4 shown) can cause this probability to change greatly around the orbit.}
	\label{figure:prob}
\end{figure}

The 10$^{32}$\,ergs\,s$^{-1}$ quiescent emission seen by {\it Suzaku} \citep{Rampy:2009tt} is very low and not easily explained by steady, spherically symmetric Bondi-Hoyle accretion \citep{1944MNRAS.104..273B} which gives a mass loss rate for the O-type star of \.M $\sim$ 10$^{-10}$ M$_{\sun}$\,yr$^{-1}$ based on a NS at $\sim$2\,R$_{\star}$.  The increase in luminosity to 10$^{36}$\,ergs\,s$^{-1}$ during a burst would require a 4 magnitude increase in the accretion rate, bringing the mass loss rate up to the more common 10$^{-6}$ M$_{\sun}$\,yr$^{-1}$ usually seen from a supergiant donor.  This change cannot be explained simply by the change between periastron and apastron.  One interpretation would be that the wind is clearly not homogeneous and isotropic. If we suggest that the bursts are caused by a clumpy wind or a similar outflow, the probability of the neutron star interacting with a clump that keeps a {\it constant} size as it moves out will be proportional to 1/r$^2$.  \citet{Ducci:2009yq} showed that the size of a clump changes as it moves away from the donor star, R$_{cl} \propto (r^{2}(1-1/r)^{\beta})^{1/3}$; where $\beta$ is a constant in the range $\sim$0.5--1.5 dependent on the supergiant and for the rest of this paper assumed to be 1.0.  This will result in the probability of interacting with a clump being proportional to $(r^{2/3}-r^{1/3})/r^2$ resulting in a change in the probability even with a slight eccentricity in the orbit (Figure~\ref{figure:prob}).  If we take the normalized probabilities under these curves between 0.8 to 0.2 at periastron and 0.3 to 0.7 at apastron we can compare these probabilities calculated from the distribution of phase folded bursts.  We observe 18 bursts at periastron and 2 at apastron out of a total of 23 bursts.  This results in relative probabilities of 0.8 and 0.1 for the phases of 0.8 to 0.2 and 0.3 to 0.7 respectively, although the low number of observed bursts, especially at apastron, will skew the probabilities in favour of more bursts being seen at periastron.  The low statistics in our sample for this letter currently prohibit any realistic attempt at testing this model.  However, with a larger sample of bursts from a large number of systems the models for a clumpy structured wind could be tested.
\section{Conclusions}
We have shown that IGR J17544--2619 has an orbital period of $\sim$4.9\,d resulting in a semi-major axis of 36-38 R$_{\sun}$.  As there is no evidence of Roche-lobe overflow, this puts an upper limit on the size of the donor star of $\sim$23R$_{\sun}$ with no eccentricity in the orbit.  However, some eccentricity is required to explain the variation of emission around the orbit and so the supergiant radius must be closer to its lower limit of 12.7R$_{\sun}$. With a larger sample of bursts and with the techniques explained here, it may be possible to better constrain the eccentricity and therefore the supergiant radius, and lead to a much better understanding of the SFXT systems seen by {\it INTEGRAL}.
\section*{Acknowledgments}
Based on observations with {\em INTEGRAL}, an ESA project funded by member states (especially the PI countries: Denmark, France, Germany, Italy, Switzerland, Spain), Czech Republic and Poland, and with the participation of Russia and the USA. A. B. Hill acknowledges support from the European Community via contract ERC-StG-200911.

\label{lastpage}

\end{document}